\documentclass[pra,twocolumn,showpacs,preprintnumbers,amsmath,amssymb,superscriptaddress,longbibliography,floatfix]{revtex4-2}

\usepackage{color}
\usepackage{graphicx}
\usepackage{dcolumn}
\usepackage{bm}
\usepackage{hyperref}
\usepackage{mathrsfs}
\usepackage{soul}
\usepackage{ulem}
\usepackage{tabularx}
\usepackage{leftidx}
\usepackage[capitalize]{cleveref}
\usepackage{enumitem}
\usepackage{booktabs}
\usepackage[T1]{fontenc} 

\setlist[itemize]{leftmargin=*}

\begin{document}


\title{Quantum repeaters with encoding on nitrogen-vacancy center platforms}

\author{Yumang Jing}
\affiliation{School of Electronic and Electrical Engineering, University of Leeds, Leeds, LS2 9JT, U.K.}

\author{Mohsen Razavi}
\affiliation{School of Electronic and Electrical Engineering, University of Leeds, Leeds, LS2 9JT, U.K.}

\begin{abstract}
We investigate quantum repeater protocols that rely on three-qubit repetition codes using nitrogen-vacancy (NV) centers in diamond as quantum memories. NV centers offer a two-qubit register, corresponding to their electron and nuclear spins, which makes it possible to perform deterministic two-qubit operations within one NV center. For quantum repeater applications, we, however, need to do joint operations on two separate NV centers. Here, we study two NV-center based repeater structures that enable such deterministic joint operations. One structure offers less consumption of classical communication, at the cost of more computation overhead, whereas the other one relies on a fewer number of physical resources and operations. We assess and compare their performance for the task of secret key generation under the influence of noise and decoherence with current and near-term experimental parameters. We quantify the regimes of operation, where one structure outperforms the other, and find the regions where encoded quantum repeaters offer practical advantages over their non-encoded counterparts.
\end{abstract}

\date{\today}
      
\maketitle

\section{Introduction}


The unavoidable transmission loss in optical channels poses a serious challenge to distributing entanglement between remote parties. A key solution to this problem is to use quantum repeaters (QRs) \cite{briegel1998,dur1999quantum}, which are the main building blocks of future quantum communications networks. The conventional idea behind QRs is to create entanglement over shorter segments, followed by entanglement swapping (ES) at all intermediate nodes to distribute entanglement over a long distance \cite{briegel1998}. Doing ES operations in a nested way can result in accumulation of errors in the system, for which entanglement distillation (ED) techniques are proposed \cite{dur1999quantum}. An attractive option to implement a repeater chain is based on using probabilistic ES operations \cite{duan2001,sangouard2011quantum}, e.g., by using linear optics, and/or probabilistic ED procedures \cite{bennett1996purification, deutsch1996quantum}.  However, due to the probabilistic nature of such operations, the finite coherence time of currently available quantum memories (QMs) can restrict the performance of such systems. Nevertheless, as quantum technologies make progress, other promising experimental platforms emerge in which at least some of the previous probabilistic steps can be operated in a deterministic way. In this paper, we focus on one such possible implementation, where both ES and ED operations are performed in a deterministic way. We particularly investigate the suitability of nitrogen-vacancy (NV) centers in diamond in such settings.

In order to perform ES and ED operations in a deterministic way, we not only need a suitable physical platform that allows deterministic two-qubit gates, but also a repeater protocol that allows one-way ED. One approach to deterministic ED operations is based on quantum repeater with encoding \cite{jiang2009quantum}. Such repeaters rely on quantum error correction (QEC) codes for their ED operations, and by doing so they go around the bottleneck caused by the transmission delays in acknowledging the success of ES and ED operations. In this protocol, the entangled states over elementary links are in encoded forms, such that errors in the ES steps can potentially be detected and corrected. That is, the ED operation effectively is performed via the QEC framework. This results in less waiting time, thus less restrictions on the QM coherence times, and can boost the entanglement generation rate in a QR. Such an approach is, however, only possible in QM platforms that allow for the deterministic gates needed for ES and QEC operations. 

In this work, we study the use of NV centers as a platform for QRs with encoding. This is partly driven by the successful implementation of deterministic two-qubit gates between electron and nuclear spins of a single NV center \cite{jelezko2004observation,waldherr2014quantum,taminiau2014universal}. Moreover, such memories are adopted for the first demonstration of a simple QR network between four cities in Netherlands \cite{van2017multiplexed,pompili2021realization}. This offers a promising platform for the implementation of  near-future encoded QR structures, as our recent work on QRs with encoding \cite{jing2020quantum,jing2020simple} suggests that the simple three-qubit repetition code could be the best option, in such repeaters, for quantum key distribution (QKD) applications over short to moderately long distances. In particular, we find that there are working regimes of operation where encoded QRs can outperform probabilistic QRs \cite{jing2020simple}. This means that for the type of networks that we are expecting to have in the short term, it could be a rewarding exercise to implement encoded QRs despite their additional implementation challenges. 

To get a sensible view of the requirements, versus gains, for NV-center based QRs with encoding, we need to consider realistic scenarios that such memories can be used in. While, in earlier works by our group \cite{jing2020quantum,jing2020simple}, the performance of the QRs with three-qubit repetition codes is carefully studied in the presence of operational errors, such analyses are not directly applicable to the case of NV centers. Firstly, the work in \cite{jing2020quantum,jing2020simple} assumes that a direct deterministic Bell-state measurement (BSM) on two separate QMs is readily available. This is not exactly the case for NV centers. While it is possible to use an entangled link between the electron spins of two NV centers to mediate a joint operation on them \cite{Kok-PRA-NV-Rep}, we should account for additional errors, or delays, that this may cause. There are also different QR structures that we can then come up with based on this mediatory entangled link, which need to be comparatively studied. Finally, the work in \cite{jing2020quantum,jing2020simple} ignores the impact of memory decoherence. Now that we have a chosen memory, which is short of ideal once it gets to coherence times, we should consider its effect on the performance to have a better assessment of system requirements.

In this work, motivated by the ideas and structures in \cite{Kok-PRA-NV-Rep} and \cite{childress2006fault}, we propose two structures for encoded QRs with NV centers. One structure has the advantage of requiring less consumption of classical communication, while the other one uses fewer resources. We will assess and compare their performance for generating secret key under the influence of erroneous operations and decoherence, using current or near-term experimental parameters. We compare the results with the simpler non-encoded structures where deterministic BSMs are employed but no ED operation is applied. Our results suggest that, while at short distances, the non-encoded schemes may offer better performance, as we go to longer distances, it pays off to use structures that employ more encoded links. {As experimental parameters, e.g., memory coherence times, improve, the encoded structure with fewer resources often offer the best performance among those considered in this work.} We also specify the gap between what we have experimentally available today versus the minimum required specifications for any of these systems to work.

The paper is structured as follows. In Sec. \ref{sec:system_description}, we begin with a description of the ideal implementation of encoded QRs motivated by Refs.~\cite{jiang2009quantum,childress2005fault} on NV-center based platforms, and explain the error models we use to formulate the problem in hand. In Sec. \ref{sec:noisy_implementation}, we analyse the effect of decoherence, as well as other system imperfections, on system performance, and calculate the secret key generation rates for such setups in Sec.~\ref{sec:system_performance}. We compare our results with the case of QRs without encoding, and illustrate the parameter regions where one type of protocol outperforms the others. Finally, we conclude the paper in Sec. \ref{sec:conclusion}.

\section{System description}
\label{sec:system_description}

In this work, we study the implementation of QRs with encoding on NV center platforms. One of the key features of NV centers, which makes them a desirable option for QR setups, is their being a two-qubit register. This includes an electron spin acting as the optical interface with single photons, and a nuclear spin, due to neighboring carbon or nitrogen atoms to the vacancy, suitable for long-time quantum storage. Moreover, using microwave and radio frequency signals, within each NV center, two-qubit operations, e.g. controlled not ({\sc cnot}) and controlled phase gates, can be performed deterministically on these two qubits \cite{wei2013compact,everitt2014high}. Within each NV center, one can also map a quantum state from the electron to the nuclear spin, and vice versa \cite{awschalom2018quantum,doherty2013nitrogen,dutt2007quantum,neumann2010single}. All these tools come handy in dealing with operations that we need in the QR setup.

An additional requirement for an efficient QR setup is the ability to write and read single photons to and from a QM. By driving an NV center, embedded in a diamond crystal, with a laser field, we can drive many transitions that mostly involve vibrational mode phonons. Such transitions will not be useful for coherent operations as these vibrational modes often quickly die out within the crystal. Zero phonon line (ZPL) emissions are then effectively the key to generating entangled states with NV centers. Even at near zero Kelvin temperatures, however, such emission are typically only a small portion, around 3\%, of all radiations from the NV center \cite{barclay2011hybrid}. Accounting also for low collection efficiency from a bulk crystal, entanglement generation with NV centers has been extremely inefficient \cite{epstein2005anisotropic}. A remedy to both problems of ZPL emission rates and collection efficiency is to have a microcavity around the NV center \cite{hausmann2013coupling,ruf2021resonant,bogdanovic2017design,nemoto2014photonic}. There have been several efforts in this regard, which have improved the ZPL emission rates to 46\% and have increased the collection efficiency by several factors \cite{riedel2017deterministic,le2012efficient,faraon2011resonant}. 
In this work, we assume cavity-based NV center platforms are available, and use known techniques with this technology to entangle light with NV centers and perform quantum operations and measurements on them. These tools are summarized in Sec.~\ref{Sec:Toolbox}, based on which, we explain several QR protocols and structures, and then finish this section with our error models. Note that the methodologies developed here are also applicable to a broader class of two-qubit memories, including silicon vacancy centers \cite{Bhaskar2020}.

Throughout the paper, we denote electron (nuclear) spins with lower (upper) case letters, for instance, if $|0\rangle_a$ and $|1\rangle_a$ represent the basis vectors corresponding to, respectively, electron spin numbers $m_S = 0$ and $m_S = -1$, then $|0\rangle_A$ and $|1\rangle_A$ represent the basis vectors corresponding to, respectively, nuclear spin numbers $m_I = 0$ and $m_I = -1$ of the same NV center. 

\subsection{NV Center as a Toolbox}
\label{Sec:Toolbox}

Here, we explain how specific features of NV centers can be used to implement the main components of encoded QRs.

\subsubsection{Entanglement distribution}
\label{Sec:entg-dist}
 One of the key ingredients of QR protocols is to establish entangled states over elementary links. Suppose we want to share an entangled state $| \Phi^+ \rangle_{AB} = \frac{1}{\sqrt{2}} (|00\rangle_{AB} + |11\rangle_{AB})$ between nuclear spins $A$ and $B$. We can then first share an entangled state between the corresponding electron spins $a$ and $b$, and then map the state of $a$ ($b$) to $A$ ($B$). This mapping is performed by initialising the nuclear spins in $|00\rangle_{AB}$, performing {\sc cnot} gates within each NV center with the electron spin as the control qubit, and then measuring the electron spins in $X$ basis.
 
 There are several schemes for distributing entangled states between the electron spins of two remote NV centers. In most of them, the entanglement distribution involves generating a spin-photon entanglement at each end of the link and then swapping entanglement in the middle of the link \cite{bernien2012two,pfaff2014unconditional, hensen2015loophole}. Depending on whether the spin-photon entanglement is in one optical mode (i.e., zero or one photon space), or two (e.g., the polarization, or time-bin, space), the Bell-state measurement (BSM) in the middle may rely on single-mode or two-mode interference. If the BSM is conclusive then the entanglement distribution task is heralded to be successful, otherwise it needs to be repeated until success. The schemes that rely on single-mode interference often require one photon to safely travel to the middle of the link, hence may have better rate scaling with distance {for heralding success}. However, in order to obtain a high fidelity entangled state, we should either keep the spin-photon entanglement generation rate very low (e.g., around 1\%) \cite{cabrillo1999creation,humphreys2018deterministic,rozpkedek2019near,pompili2021realization}
 or rerun the procedure to distill the entangled state \cite{barrett2005efficient,pfaff2014unconditional,bernien2013heralded, nemoto2014photonic}. {In both cases the effective success rate, in certain regimes of interest to this work, could then become comparable to the two-mode schemes, where the rate decays exponentially with the distance between nodes $A$ and $B$. For instance, in our setup, where the optimum elementary link is typically below 20~km, the extra channel loss in the two-mode case is a small factor, although one has to also account for additional coupling or detector efficiencies. But, aside from the success rate of entanglement generation, another important factor is the amount of the initial noise, or loss in fidelity, that we can tolerate in our system. The two-mode schemes can, in principle, generate ideal entangled states, whereas, in the single-mode schemes, some errors, due to, e.g., generating one photon at each end, would be inevitable. In a real experiment, one has to factor in all these nuances, as well as practical restrictions on the system, to decide which entanglement distribution scheme may work best in their setting. }
 
 In order to encompass the essence of different entanglement distribution schemes available, here, we assume that a generic two-mode entanglement distribution scheme is used where, at each of nodes $A$ and $B$, the polarization of a single photon is entangled with the electron spin of the NV center (see \cite{Nicolo-NV1, Nicolo-NV2}, for example). These photons are frequency converted, if needed, and will be coupled to an optical channel. Using linear optics and single-photon detectors, a partial BSM in the polarization basis is then performed on these two photons at the middle of the link. Once a successful BSM is heralded, this information will be sent back to nodes $A$ and $B$, at which point, the state of electron spins is transferred and stored onto the corresponding nuclear spins. We model the generated entangled state as a Werner state as will be explained later in the next section. {Note that any other entanglement distribution scheme can also be similarly analyzed using techniques and procedures provided in this work.}

\subsubsection{Encoded entanglement distribution}
\label{Sec:enc-entg}

In this work, we consider encoded QRs with three-qubit repetition codes, where the logical qubits are encoded as
\begin{align}
    |\tilde{0} \rangle = |000\rangle \quad \text{and} \quad  |\tilde{1} \rangle = |111\rangle.
\end{align}
where $|0\rangle$ and $|1\rangle$ represent the standard basis for a single qubit. This code can correct up to one bit-flip error. Although this is not a strong error correction code, it has been shown in our previous work that so long as we rely on its error detection features it offers a reasonable performance at short and moderately long distances \cite{jing2020quantum, jing2020simple} as compared to more complicated codes. We therefore analyse this particular code  for our NV center platform.

\begin{figure}[t]
\includegraphics[width=0.8\columnwidth]{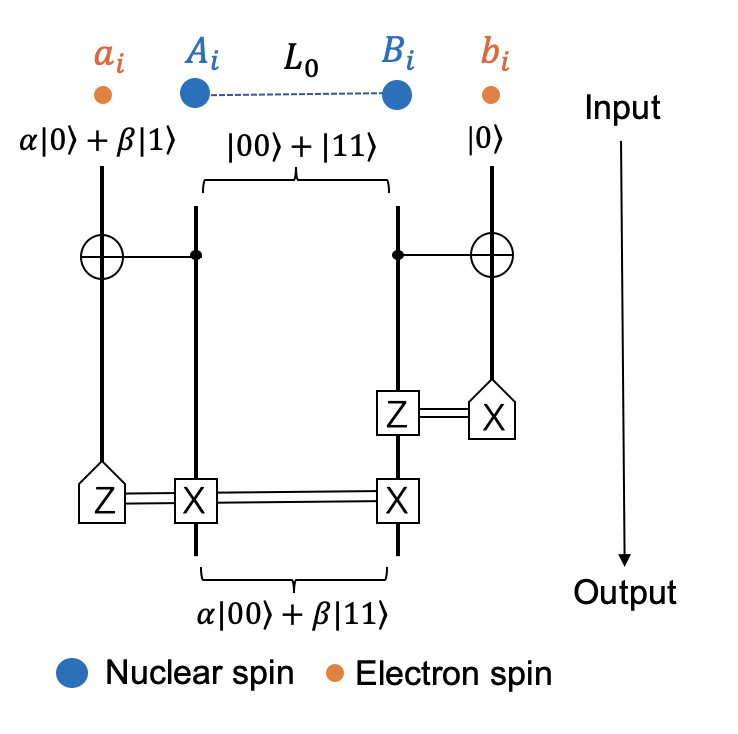}
\caption{\label{remote_cnot} Quantum circuit for remote {\sc cnot} gate. Note that single-qubit measurements (trapezoidal boxes) are performed on electron spins. Here, $a_i-A_i$ represent the electron-nuclear spins in one NV center separated by a distance $L_0$ from the corresponding NV center, $b_i-B_i$, at the other end of the elementary link. 
}
\end{figure}

The first step in an encoded QR is to ideally distribute encoded entangled states in the following form:
\begin{align}
    |\tilde{\Phi}^+\rangle_{\mathbf{AB}}=\frac{1}{\sqrt{2}} (|\tilde{0}\tilde{0}\rangle_{\mathbf{AB}} +|\tilde{1}\tilde{1}\rangle_{\mathbf{AB}}),
    \label{phi+}
\end{align}
where here we consider two example memory banks $\mathbf{A}=(A_1,A_2,A_3)$ and $\mathbf{B}=(B_1,B_2,B_3)$ at the two ends of an elementary link. To this end, using the scheme described in the previous subsection, we first generate Bell pairs $| \Phi^+ \rangle_{A_iB_i} = \frac{1}{\sqrt{2}} (|00\rangle_{A_iB_i} + |11\rangle_{A_iB_i})$, for $i=1,2,3$. Once electron spins are available again in all NV centers of memory banks $\mathbf{a}$ and $\mathbf{b}$, we initialize them in the codeword states $\frac{1}{\sqrt{2}} (|\tilde{0}\rangle_\mathbf{a} +|\tilde{1}\rangle_\mathbf{a})$ and $|\tilde{0}\rangle_\mathbf{b}$. Finally, using transversal remote {\sc cnot} gates, shown in Fig.~\ref{remote_cnot}, we can generate the state in Eq.~\eqref{phi+} \cite{jiang2009quantum}. The same procedure is applied to all elementary links.

Note that the remote {\sc cnot} circuit in Fig.~\ref{remote_cnot} is slightly different from the one used in \cite{jing2020quantum, jing2020simple}. In the latter work, the remote {\sc cnot} circuit requires measurements on {qubits that hold the initial Bell state, i.e., the nuclear spins in our NV-center setup}. In NV centers, however, a nuclear spin is often measured by first mapping its state to an electron spin, using a {\sc cnot} gate, and then measuring the electron spin  \cite{neumann2008multipartite,neumann2010single}. This is not, however, possible in our case as this would ruin the initial state of the electron spins. We have therefore slightly changed the remote {\sc cnot} circuit such that the measurements are only done on electron spins with nuclear spins always in an entangled state.

\subsubsection{Entanglement Swapping}
\label{Sec:entg-swap}

Once encoded entangled states are distributed between the nuclear spins across all elementary links, the next step is to perform entanglement swapping (ES) operations at all intermediate stations to extend the entanglement to the entire link. In the encoded repeater protocol, this can be done by performing BSMs, in a transversal way, on corresponding pairs of NV centers at each of the intermediate nodes. {This operation would also allow us to pick up some of the errors that might have been accumulated by this stage, and help us distill the final entangled state. {For instance, in the 3-qubit repetition code considered here, the BSM is made of an $X$ and a $Z$ operator measurement, the results of which specify the type of encoded Bell state that will be shared between the remote nodes. Ideally, the results of the $Z$ operator measurements must be $000$ or $111$. Because of the errors in the system, we may, however, get other combinations of $0$ and $1$, which correspond to detecting an error. The majority rule here can be used to specify the most likely post-BSM encoded Bell state.} It turns out \cite{jing2020quantum}, however, that for QKD purposes, detecting the error, and using that information for post-selection, would provide us with an effective way to boost the key rate, and error correction, as envisaged in the original protocol \cite{jiang2009quantum}, may not be needed.}

For the above process, a direct joint measurement on the nuclear spins of two separate, although possibly co-located, NV centers may not be possible. To do a deterministic BSM on two separate nuclear spins, here, we distribute an additional Bell pair between the corresponding electron spins of the two NV centers. ES operations can then be performed by performing BSMs on the nuclear and electron spins within each NV center \cite{Kok-PRA-NV-Rep}. This can be done by first applying a {\sc cnot} gate to nuclear and electron spins followed by relevant single-qubit measurements on each. Note that, in this procedure, we have to first measure the electron spin, and then map the nuclear spin state to the electron spin. The latter can be done by initialising the electron spin in an appropriate state and then performing a {\sc cnot} gate on the two spins with the nuclear spin as the control qubit. We can then measure the electron spin again, to effectively complete the measurement on the nuclear spin. A similar procedure can be used across the repeater chain. The measurement outcomes need to be notified to the end users to adjust the Pauli-frame on the final states, and/or for error correction or post-selection purposes. 

Note that in the above procedure, the two NV centers do not necessarily need to be co-located, and, in principle, one can assume an arbitrary distance between the two memories. That would, however, change system resilience to memory decoherence. To study this, in the following, we define several protocols for different QR architectures and will analyse and compare them in the forthcoming sections.

\subsection{Quantum repeater structures and protocols}
In this section, based on whether we employ coding or not, and how deterministic BSMs are done, we define four protocols, as explained below.

\subsubsection{Protocols for encoded repeaters}

\begin{figure}[tbh]
\includegraphics[width=0.9\columnwidth]{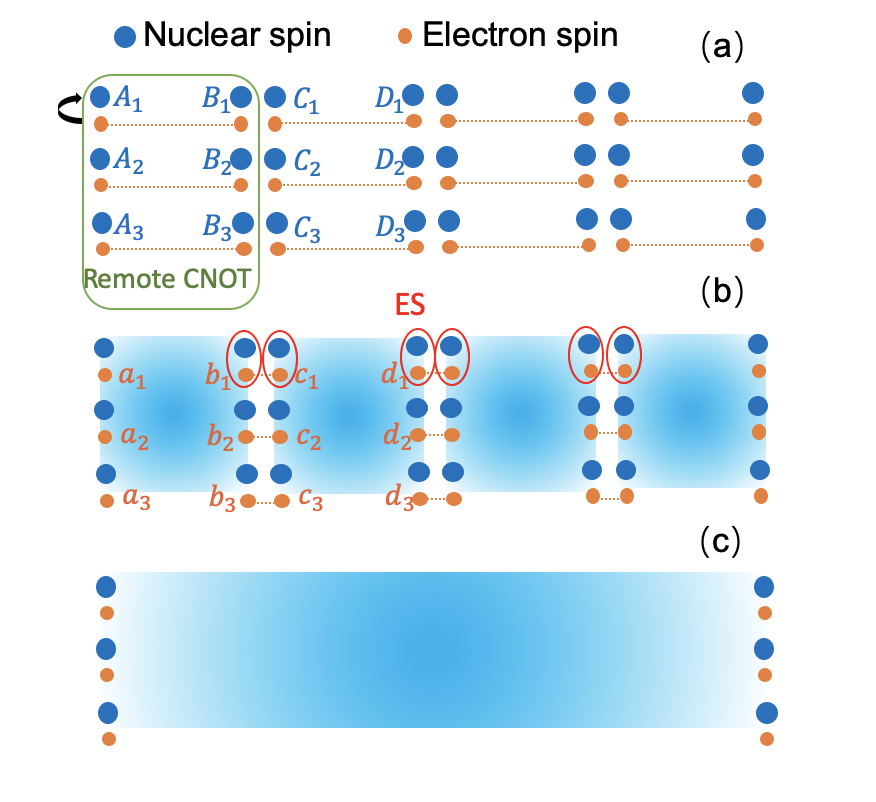}
\caption{\label{schematic_1} Schematic QR structure for protocol 1 with the following steps: (a) Distributing Bell pairs between electron spins (small orange ovals) over all elementary links in a heralding way. Transferring and storing the entangled states to the corresponding nuclear spins (large blue ovals), followed by remote {\sc cnot} gate. (b) Performing ES operation on nuclear spins at intermediate nodes by creating temporary Bell pairs between the corresponding electron spins, and then performing a BSM within each NV center. (c) The final encoded entangled state is created between the two end users. Based on the measurement results at each middle node, the Pauli frame of the final entangled state can be adjusted.}
\end{figure}

\begin{figure}[t]
\includegraphics[width=0.9\columnwidth]{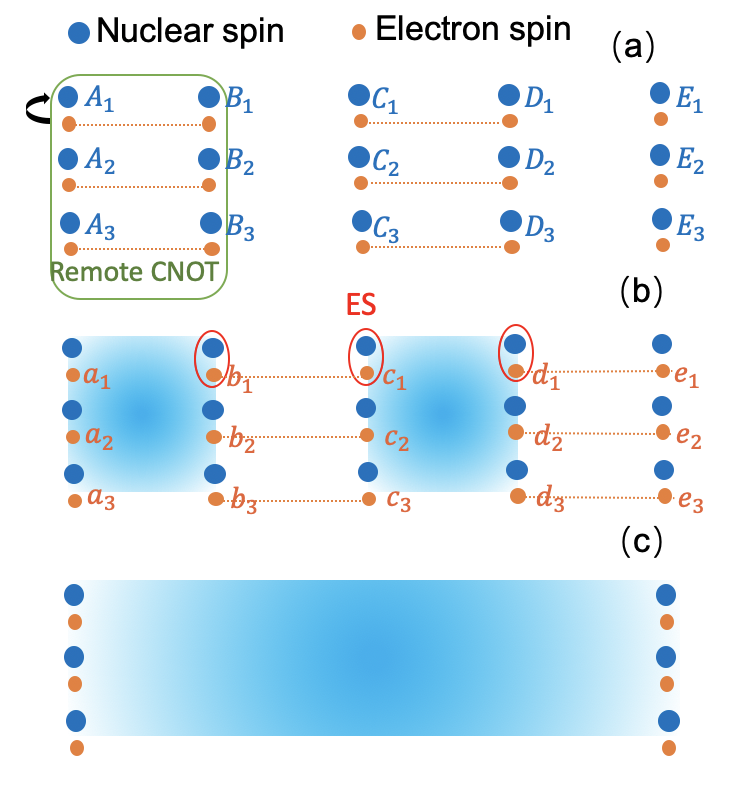}
\caption{\label{schematic_2} Schematic QR structure for protocol 2 with the following steps: (a) Generating encoded Bell pairs between nuclear spins in every other link; (b) Distributing Bell pairs between electron spins in all remaining links in order to facilitate BSM within each NV center at intermediate nodes. (c) The encoded entanglement is extended to end users. Based on the measurement outcomes gathered from middle stations, one can adjust the Pauli frame of the final entangled state. }
\end{figure}

Here, we describe the ideal implementation of the protocol proposed in Ref. \cite{jiang2009quantum} with three-qubit repetition codes on NV center platforms. We consider two architectures, shown in Figs.~\ref{schematic_1} and \ref{schematic_2}, depending on whether the BSM is done on co-located NV centers or those apart by a distance $L_0$, corresponding to the length of an elementary link. In both structures, there are a total of $2^n$ elementary links, where $n$ is the nesting level of the corresponding QR.

In what we refer to as protocol 1 (P1), we use the structure in Fig.~\ref{schematic_1}, and carry out the following steps:  
\begin{itemize}
  \item Step 1: distribute encoded entanglement across all elementary links; see Sec.~\ref{Sec:enc-entg}. {As this requires multiple attempts to entangle all relevant pairs of NV centers, we stop this process, whether or not all relevant pairs are entangled, after a stoppage time $T_1$ and move to the next step.}
  \item Step 2: perform BSMs at all intermediate nodes; see Sec.~\ref{Sec:entg-swap}. {We again stop this procedure, whether or not all relevant BSMs are completed, after a stoppage time $T_2$.}
  \item Step 3: pass all measurement results to the two end users. {If there are missing entangled pairs, or incomplete BSMs, then we discard the state generated in that round. We will account for the effect of such discarded states in our key rate analysis.}
\end{itemize}

In what we call protocol 2 (P2), motivated by Refs.~\cite{childress2005fault,childress2006fault}, we use the structure in Fig.~\ref{schematic_2}, and carry out the following steps:  
\begin{itemize}
  \item Step 1: distribute encoded entanglement across every other elementary link; see Sec.~\ref{Sec:enc-entg}. {We stop this process after a stoppage time $T_1$ and move to the next step.}
  \item Step 2: distribute uncoded entanglement across the electron spins of all remaining links; see Sec.~\ref{Sec:entg-dist}. {We stop this process after a stoppage time $T_2$ and move to the next step.}
  \item Step 3: perform BSMs at all intermediate nodes, which now only contain single NV centers; see Sec.~\ref{Sec:entg-swap}.
  \item Step 4: pass all measurement results to the two end users. {If there are missing entangled pairs, or incomplete BSMs, then we discard the state generated in that round.}
\end{itemize}
Protocol 2 requires fewer NV centers and operations than protocol 1, and, in that sense, may offer some advantage. But, in the end, what matters is the overall performance, normalized by the total number of memories used, which we use for comparison between such protocols. 

\subsubsection{Protocols for uncoded repeaters}
In order to better understand whether QRs with encoding will offer any advantages over their non-encoded versions, in this work, we also consider protocols 3 (P3) and 4 (P4), which are, respectively, the uncoded versions of protocols 1 and 2. For simplicity, we do not consider any distillations for protocols 3 and 4, as, without coding, that would turn them into probabilistic protocols. A comparison between encoded QRs and probabilistic ones is already available in \cite{jing2020simple}. In protocols 3 and 4, we just need to replace Step 1 with the following revised step:
\begin{itemize}
\item Step 1': distribute Bell pairs between nuclear spins in all, for protocol 3, or every other, for protocol 4, elementary links; see Sec.~\ref{Sec:entg-dist}. {We move to the next step after a stoppage time $T_1$.}
\end{itemize}
The remaining steps are as in protocols 1 and 2, respectively.

\subsection{Error models}
\label{subsec:error_model}

In order to analyse the above QR setups, we consider three major sources of imperfections as follows.

\noindent
(1) \textbf{Gate imperfections:} The {\sc cnot} gate for a control nuclear spin $J$ and a target electron spin $j$, within an NV center, is modeled as ~\cite{briegel1998}
\begin{align}
\label{eq:CNOTmodel}
\rho^{\text{out}} = (1-\beta) U_{J,j} \rho^{\text{in}} U_{J,j}^{\dagger} + \frac{\beta}{4} \text{Tr}_{J,j} (\rho^{\text{in}}) \otimes \mathbb{I}_{J,j},
\end{align} 
where $\rho^{\text{in}}$ ($\rho^{\text{out}}$) is the input (output) before (after) the {\sc cnot} gate, and $U_{J,j}$ represents the unitary operator corresponding to an ideal {\sc cnot} gate. The error in this two-qubit operation is modeled by a uniform depolarization of qubits $J$ and $j$, represented by identity operator $\mathbb{I}_{J,j}$, with probability $\beta$. We assume that a similar relationship as in Eq.~\eqref{eq:CNOTmodel} would also model a {\sc cnot} gate with the electron (nuclear) spin as the control (target) qubit. While not necessarily the case, for simplicity, we assume that the parameter $\beta$ is the same in both cases. In NV centers, there are other common two-qubit gates, such as controlled phase gates, that may be used in practice. Using equivalent quantum circuits, however, such operations can often be modeled by a {\sc cnot} gate with possibly additional single-qubit rotations. In such cases, we assume parameter $\beta$ captures the total error in the equivalent model. As in Ref. \cite{jing2020quantum,jing2020simple}, here, we assume all single-qubit operations are perfect.

\noindent
(2) \textbf{Measurement errors:} The projective measurements to electron spin states $|0\rangle$ and $|1\rangle$ are, respectively, represented by 
\begin{align}
\label{eq:meas}
P_0 &= (1-\delta) |0\rangle \langle 0| + \delta |1\rangle \langle 1| \quad \mbox{and} \nonumber\\
P_1 &= (1-\delta) |1\rangle \langle 1| + \delta |0\rangle \langle 0|,
\end{align}
where $\delta$ is the measurement error probability. Similar measurement operators, $P_\pm$, are used for projective measurement in $|\pm \rangle = 1/\sqrt{2}(|0 \rangle \pm |1 \rangle)$ basis. The projective measurements of nuclear spins are modelled effectively in the same format but with error parameters {$\beta/2+\delta$}, since there should always be a mapping operation performed through a {\sc cnot} gate as described in Sec. \ref{Sec:entg-swap}.

\noindent
(3) \textbf{Decoherence:} We model the decoherence effect in electron/nuclear spins by using a depolarizing channel. For a single qubit $a$ ($A$), after a waiting time $t_{\rm w}$, the initial state $\rho$ will be mapped to
\begin{align}
  &  \mathcal{D}_{\rm depol}^{a}(\rho)=\lambda_2^{\rm e} \rho + (1-\lambda_2^{\rm e}) (\mathbb{I}_2 - \rho), \nonumber \\
  &  \mathcal{D}_{\rm depol}^{A}(\rho)=\lambda_2^{\rm n} \rho + (1-\lambda_2^{\rm n}) (\mathbb{I}_2 - \rho),
\label{eq:depol_model}
\end{align}
where
\begin{align}
    \lambda_2^{\rm e/n} (t_{\rm w})= \frac{1}{2} + \frac{e^{-\frac{t_{\rm w}}{\tau_{\rm e/n}}}}{2}
    \label{eq:lam2}
\end{align}
with $\tau_{\rm e/n}$ being the coherence time for electron/nuclear spins and $\mathbb{I}_d$ being a $d \times d$ identity matrix. The expression in Eq.~\eqref{eq:depol_model} is a re-arranged form of the typical expression for a depolarizing channel, $p \rho + (1-p) \mathbb{I}_2 /2$, with $p = \exp{(-{t_{\rm w}}/{\tau_{\rm e/n}})}$, in which $\lambda_2^{\rm e/n}$ represents the fidelity of the output state with respect to {the input state, in the case of pure input states}. As shown below, this formulation suits better the two special cases of interest we need to deal with in the setup under consideration.

The first case of interest is when we have a two-qubit system in an initial entangled state such as $|\Phi^+\rangle_{AB (ab)}$, for nuclear spins (electron spins). After a waiting time $t_{\rm w}$, both spins decohere according to Eq.~\eqref{eq:depol_model} resulting in $\mathcal{D}_{\rm depol}^{AB} (|\Phi^+\rangle_{AB} \langle\Phi^+|)$ as the output state for nuclear spins, where $\mathcal{D}_{\rm depol}^{AB}=\mathcal{D}_{\rm depol}^{A} \circ \mathcal{D}_{\rm depol}^{B}$, and similarly for electron spins. As shown in Appendix \ref{sec:appendix}, for $\rho = |\Phi^+\rangle_{AB (ab)} \langle\Phi^+|$, the output state can be written as
\begin{align}
    &\mathcal{D}_{\rm depol}^{AB} (\rho)=\lambda_4^{\rm n} \rho + (1-\lambda_4^{\rm n}) (\mathbb{I}_4 - \rho)/3, \nonumber \\
    & \mathcal{D}_{\rm depol}^{ab} (\rho)=\lambda_4^{\rm e} \rho + (1-\lambda_4^{\rm e}) (\mathbb{I}_4 - \rho)/3,
\label{eq:depol_model2}
\end{align}
where 
\begin{align}
    \lambda_4^{\rm e/n} (t_{\rm w}) = \frac{1}{4} (3\lambda_2^{\rm e/n } (t_{\rm w})-1)^2 + \frac{3}{4} (1-\lambda_2^{\rm e/n}(t_{\rm w}))^2
\label{eq:lambda4}
\end{align}
is the fidelity of the output state with respect to the entangled input state. The same form as in  Eq.~\eqref{eq:depol_model2} holds for any other Bell states, or any mixed state diagonal in Bell states, such as Werner states. For a general two-qubit state, Eq.~\eqref{eq:depol_model2} acts as a conservative approximation to decoherence effects given that it correctly specifies the fidelity of the output state, while maximizing the noise by using a maximally mixed state for all other off-diagonal terms. We use \cref{eq:depol_model2} to model decoherence across the elementary links as, in practical regions of interest, deviations from a Bell-diagonal state is reasonably small. Note that the off-diagonal terms ignored by our approximation often do not contribute to quantum bit error rate (QBER) in QKD systems.

The second case of interest is when the initial state is of the form $|\tilde{\Phi}^+\rangle_{\mathbf{AB}}$, which is a six-qubit system, or a slight deviation from it. With similar calculations, we approximate the output state for an encoded entangled state $\rho$ by
\begin{align}
    \mathcal{D}_{\rm depol}^{\mathbf{AB}}(\rho)=\lambda_{64}^{\rm n} \rho + (1-\lambda_{64}^{\rm n}) (\mathbb{I}_{64} - \rho)/63,
\label{eq:depol_model6}
\end{align}
where 
\begin{align}
    \lambda_{64}^{\rm e/n} (t_{\rm w})= &\frac{1}{64} [(3\lambda_2^{\rm e/n} (t_{\rm w})-1)^6 +33 (1-\lambda_2^{\rm e/n}(t_{\rm w}))^6 \nonumber\\ &+15(3\lambda_2^{\rm e/n}(t_{\rm w})-1)^2(1-\lambda_2^{\rm e/n}(t_{\rm w}))^4 \nonumber\\
    &+ 15(3\lambda_2^{\rm e/n}(t_{\rm w})-1)^4(1-\lambda_2^{\rm e/n}(t_{\rm w}))^2]
\label{eq:lambda64}
\end{align}
is the fidelity of the output state, with respect to the input state, if the initial state is the ideal encoded entangled state $|\tilde{\Phi}^+\rangle_{\mathbf{AB}}$. Similar to the two-qubit case, the above modeling of decoherence effectively treats all non-desired states as a {maximally} mixed state while correctly predicting the output fidelity; see Appendix \ref{sec:appendix} for more detail.

\section{Error Analysis}
\label{sec:noisy_implementation}

In order to assess how well the NV-center based encoded QRs would operate, here, we obtain the final distributed state as a function of system parameters. In the case of measurement or gate errors, we have previously devised analytical and numerical techniques to accurately account for such issues and their impact on system performance \cite{jing2020quantum, jing2020simple}. In this work, we additionally account for the effect of memory decoherence especially because, in terms of coherence time, the electron spins in the NV centers may impose some limitations on the achievable rates and distance.

Accounting for the decoherence effect, in an analytical way, in a system with many individual NV centers, where each decoheres on its own independently of others, is by no means an easy task. Here, we devise an approximation technique, in which, at each step of the way, we calculate the average waiting time for memories involved, and then assume all of them have decohered by the same average time. This should provide us with a reasonable approximation to what in practice can be achieved, which is what we are looking for here in the context of QKD as an application.


In the following, we calculate the relevant time parameters for each step of the proposed protocols and explain our methodology to obtain the QR final state as a function of system parameters.

\subsection{Entanglement distribution}

Here, we first obtain the entangled state distributed over an elementary link. This involves two steps: first, generating an entangled state between two electron spins, and, then, transferring that state to the corresponding nuclear spins. In both processes, we deviate from an ideal Bell pair because of gate errors and decoherence. We follow the two-photon protocol described in Sec.~\ref{Sec:entg-dist}. For simplicity, we assume that the generated entangled state without any decoherence is the ideal Bell pair $|\Phi^+\rangle$. By the time that we hear about the success of entanglement distribution, this ideal state of electron spins has already decohered by the time it takes to transmit photons and learn about the success of the entanglement distribution protocol. In this work, we assume that, compared to the transmission time (which for our setup is typically on the order of tens of $\mu$s, or longer), the time it takes for any local operation is negligible. In that case, this waiting time, or, effectively, the repetition period for the entanglement distribution protocol is given by
\begin{align}
    T_0  = \frac{L_0 }{c},
\label{eq:T0_P1}
\end{align}
where $L_0=L_{\rm tot}/2^n$ is the length of elementary links, with $L_{\rm tot}$ being the total distance between the two end users and $n$ is the nesting level. During this time, the desired target state $|\Phi^+\rangle$ decoheres in electron spins, according to Eq.~\eqref{eq:depol_model2}, yielding
\begin{align}
    \rho_{\rm ee}= F_0 |\Phi^+\rangle \langle\Phi^+ | + \frac{1-F_0}{3} (\mathbb{I}_4-|\Phi^+\rangle \langle\Phi^+ |), 
\label{original_state}
\end{align}
which is a Werner state with 
\begin{align}
    F_0=\lambda_4^{\rm e}(T_0),
\label{eq:F0_P1}
\end{align}
where $ \lambda_4^{\rm e}$ is given by \cref{eq:lambda4}. 

This state is then immediately transferred onto the corresponding nuclear spins. This is being done by applying one {\sc cnot} gate on each end, with electron spins as the control qubit and nuclear spins in an initial state $|0\rangle$, followed by $X$ measurements on electron spins. This process has been analytically simulated, according to Eqs.~\eqref{eq:CNOTmodel} and \eqref{eq:meas}, by the symbolic software Mathematica to give us the entangled state $\rho_{\rm nn}$ shared between two nuclear spins at distance $L_0$. 

\subsection{Encoded entanglement distribution}
The next step in encoded protocols is to create encoded entanglement across certain elementary links. In principle, once the three Bell pairs required in each leg are established, we can proceed with the remote {\sc cnot} gate operation that distributes encoded entanglement across the corresponding link. In our proposed protocols, we, however, wait for a time $T_1$ before we proceed to the ES stage. This means that the nuclear spins in our system have decohered for an average time of $\overline T_1= T_1 - T_0/P_0(L_0)$, with 
\begin{align}
P_0 (L_0)=\frac{1}{2} \eta_c^2 \eta_t^2 \eta_d^2 
\label{eq:prob_ele}  
\end{align}
being the success probability for each entangling attempt, where $\eta_c$ accounts for the emission probability of a ZPL photon from the NV center, its collection and coupling efficiency into and out of the optical channel, and the efficiency of any required frequency conversion, $\eta_d$ is the single-photon detector efficiency, and $\eta_t=\text{exp}[-L_0/(2L_{\rm att})]$ is the transmissivity of a photon through half of the elementary link. Note that, per elementary link, $1/P_0 (L_0)$ is the average number of times that we have to repeat our entanglement distribution scheme until it succeeds. We have to repeat this process, in parallel, for $M$ different pairs of memories, where, in protocol 1, $M = 3 \times 2^n$, in protocol 2, $M = 3 \times 2^{n-1}$, in protocol 3, $M = 2^n$, and in protocol 4, $M = 2^{n-1}$. Here, the waiting time for each memory pair is by itself a random variable, independent of, but identically distributed with, other waiting times, for all of which the average waiting time is given by $\overline T_1$. Even if we define a statistical average waiting time variable, $T_w =(1/M) \sum_{i=1}^M T_w^{(i)}$, where $T_w^{(i)}$ is the waiting time for the $i$th link, the expected value of $T_w$ is equal to $\overline T_1$, and its variance is expected to be small for large values of $M$, as it is often the case for structures of interest in our work. In short, $\overline T_1$ properly captures the average decoherence time in this phase of our setup. As we will see later, $\overline T_1$ is indirectly a function of $M$ as our choice of $T_1$ would depend on $M$.

Based on our average approach to accounting for decoherence across the repeater chain, here we assume all nuclear spins have decohered for a time {$\overline T_1$} by the time we apply the remote {\sc cnot} gate operation for encoded repeaters. This effect can be modelled by \cref{eq:depol_model2} at $ \lambda_4^{\rm n}(\overline T_1) $, with input state $\rho = \rho_{\rm nn}$. We then model the operations in the remote {\sc cnot} circuit in Fig.~\ref{remote_cnot}, accounting for operation and measurement errors  \cite{jing2020quantum,jing2020simple}, to obtain $\rho'_{\rm nn}$ as the output state for this stage of the protocol. Note that, for remote {\sc cnot} operation, electron spins are initialized into the codeword states. {This can be done, e.g., using techniques introduced in \cite{GHZprep1, GHZprep2, GHZprep3}. Based on these techniques, in this work, we assume that the codeword states are created error-free and the time it takes to prepare them is embedded into $T_1$. This is because the remote {\sc cnot} operation can be done at each elementary link once the three required Bell states for that link are generated. That implies that, in terms of timing, the additional delay caused by the remote {\sc cnot} procedure, including the local preparation of the initial codeword states, would only matter for the elementary link that gets entangled the last. Given that $T_1$, in typical regimes of operation, is on the order of $m$s, and local operations are assumed to be much faster, we neglect this additional timing parameter. If this is not the case in a certain experiment, the parameter $T_1$ can be adjusted accordingly for rate calculations.} In protocols 3 and 4, we follow the same procedure but we do not include the remote {\sc cnot} operation. 

\subsection{Entanglement swapping}

Once encoded/uncoded entanglement is stored in the nuclear spins, additional electron-electron entanglement is established so that ES operations can be performed at intermediate stations. For protocols 2 and 4, this process is the same as what has been done for the distribution of original Bell pairs, whereas, in protocols 1 and 3, the Bell pairs are distributed only over a very short distance between two co-located electron spins. In the latter case, we assume that the corresponding electron-spin decoherence happens over a negligible time, whereas in the former the electron-electron state has the same form as $\rho_{\rm ee}$ in \cref{original_state}. 

Once electron-electron entanglement is established, the corresponding ES operations between electron and nuclear spins are immediately performed. These ES operations could therefore be performed at different times for different memories. To estimate the decoherence during this step, and to follow the simple scheme we have adopted for decoherence analysis, we calculate the average time $\overline T_2(P_s) = T_s/ P_s$, to do ES operations across the repeater chain, where $T_s$ denotes the repetition period for the electron-electron entangling attempt, and $P_s$ denotes its success probability. During this time, our state $\rho'_{\rm nn}$ would decohere. For protocol 1, the decoherence is modeled by \cref{eq:depol_model6} with $\lambda_{64}^{\rm n}$ calculated at $t_{\rm w} = \overline T_2(P_0(0))$ and $T_s$ being a small internal time constant. For protocol 2, the decoherence is modeled by \cref{eq:depol_model6} with $\lambda_{64}^{\rm n}$ calculated at  $t_{\rm w} = \overline T_2(P_0(L_0))$ and $T_s = T_0$. For protocol 3, the decoherence is modeled by \cref{eq:depol_model2} with $\lambda_{4}^{\rm n}$ calculated at $t_{\rm w} = \overline T_2(P_0(0))$ and $T_s$ being a small internal time constant. Finally, for protocol 4, the decoherence is modeled by \cref{eq:depol_model2} with $\lambda_{4}^{\rm n}$ calculated at $t_{\rm w} = \overline T_2(P_0(L_0))$ and $T_s = T_0$. 


Let us denote the resulting state after the above decoherence process as $\rho''_{\rm nn}$. Using the error models in Sec.~\ref{subsec:error_model}, and the techniques introduced in \cite{jing2020simple, jing2020quantum}, we can then calculate the final output state of the QR accounting for gate, measurement, and decoherence errors. 
 
\section{QKD Performance}
\label{sec:system_performance}

It would be interesting to compare different QR structures and protocols in terms of their performance for a concrete application. Here, we choose QKD as our benchmarking tool. We use the decoder modules proposed in \cite{jing2020simple}, which only rely on single-qubit measurements, to generate a raw key bit. We also use the post-selection technique proposed in \cite{jing2020quantum}, wherein only data points that no errors has been detected in the ES stage are used for key generation. References \cite{jing2020quantum} and \cite{jing2020simple} offer a detailed description of the analytical-numerical methods used to calculate the shared state between the two end users, from which secret key rates has been calculated in this work.  

\begin{table}[tb]
    \centering
    {
    \begin{tabular}{c|c|c}
         & $\overline T_1$, $P_{S1}$ & $\overline T_2$, $P_{S2}$ \\ \hline
       Protocol 1  & $M=3\times2^n$,  &$M=3\times(2^n-1)$, \\
         & $q=P_0(L_0)$ &$q=P_0(0)$, $T_s$ fixed \\ \hline
          Protocol 2  & $M=3\times2^{n-1}$,&$M=3\times2^{n-1}$,   \\ 
            & $q=P_0(L_0)$  &$q=P_0(L_0)$, $T_s=T_0$  \\ \hline
          Protocol 3  & $M=2^n$,  &$M=2^n-1$,  \\
            &$q=P_0(L_0)$ &$q=P_0(0)$, $T_s$ fixed \\\hline
         Protocol 4  & $M=3\times2^{n-1}$,  & $M=2^{n-1}$,  \\
           & $q=P_0(L_0)$ & $q=P_0(L_0)$, $T_s=T_0$ \\
    \end{tabular}
    \caption{The relevant values of $M$, $q$, and $T_s$ for calculating $\overline T_1$, $\overline T_2$, $P_{S1}$, and $P_{S2}$, for different protocols. \label{Tab:Mparam}} }
    
\end{table}

Here, we first calculate the secret key generation rate per entangled state between Alice and Bob {for the BBM92 protocol \cite{bennett1992quantum}}. In the asymptotic limit, and, for the efficient \cite{lo2005efficient} entanglement-based QKD protocol, where one basis is used more often than the other, this parameter, known as the secret fraction \cite{scarani2009security}, is given by 
\begin{align}
    r_{\infty}= \text{Max} \{0, 1-h(e_z)-h(e_x) \},
\end{align}
where $h(p)=-p \text{log}_2 (p) - (1-p) \text{log}_2 (1-p)$ is the Shannon binary entropy function and $e_i$ is the quantum bit error rate (QBER) in measurement basis $i$, i.e., the probability that Alice and Bob get discordant measurement outcomes in that basis. Here, the secret fraction is calculated for the better of two decoders proposed in \cite{jing2020simple}. 

In order to obtain the total secret key generation rate, we need to multiply the secret fraction by the entanglement generation rate $R$. Due to our assumptions that the time for local operations and measurements is negligible, the overall timescale for the implementation of protocols is determined by the sum of $T_1$ and $T_2$. Thus, the rate to obtain a $L_{\rm tot}$-distant entangled pair is expressed as {
\begin{align}
    R=\frac{P_S}{T_1 + T_2} , 
\label{eq:ent_generation_rate}
\end{align}
where $P_S = P_{S1}P_{S2}$ denotes the probability that, in step 1, all required elementary links are successfully entangled {\it and}, in step 2, all relevant BSMs are performed. The success probability for step 1 is given by $P_{S1} = (1-(1-q)^{T_1/T_0})^M$, and, in step 2, conditioned on success in step 1, by $P_{S2} = (1-(1-q)^{T_2/T_s})^M$, where, in each protocol, the corresponding values for $q$, $M$, and $T_s$ are outlined in \cref{Tab:Mparam}.}  In this work, we normalize secret key rate by the number of NV centers to assess and compare the performance of proposed protocols. The normalized key rate is given by
\begin{align}
    R_{\text{QKD}}^{\rm P1} &= \frac{R^{\rm P1} r_{\infty}^{\rm P1}}{6 \times 2^n}, \nonumber\\
    R_{\text{QKD}}^{\rm P2} &= \frac{R^{\rm P2} r_{\infty}^{\rm P2}}{3 \times (2^n+1)}, \nonumber\\
     R_{\text{QKD}}^{\rm P3} &= \frac{R^{\rm P3} r_{\infty}^{\rm P3}}{2^{n+1}}, \nonumber\\
     R_{\text{QKD}}^{\rm P4} &= \frac{R^{\rm P4} r_{\infty}^{\rm P4}}{2^n+1},
\label{eq:norm_skr}
\end{align}
for protocols 1 to 4, respectively, as specified by the superscripts.

\begin{figure}[thb]
\includegraphics[width=\columnwidth]{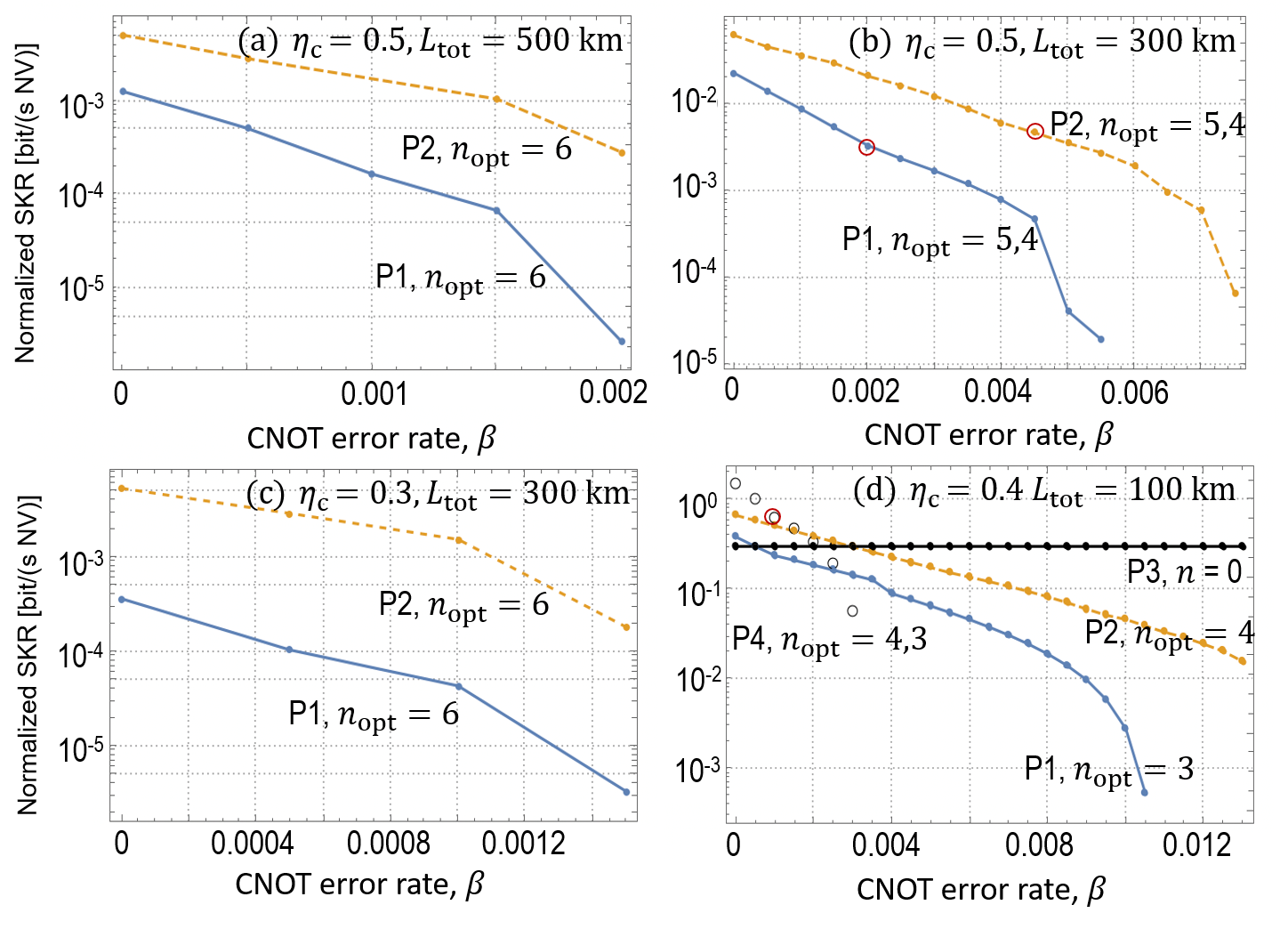}
\caption{\label{fig:SKR_beta} Comparison of normalized secret key rate as a function of {\sc cnot} error probability $\beta$ at (a) $\eta_c=0.5, L_{\rm tot}=500$~km, (b) $\eta_c=0.5, L_{\rm tot}=300$~km, (c) $\eta_c=0.3, L_{\rm tot}=300$~km, and (d) $\eta_c=0.4, L_{\rm tot}=100$~km, for protocols 1--4. The result for an entanglement distribution link without encoding, i.e., effectively, protocol 3 at $n=0$, is also calculated (black solid curve in (d)). The optimum nesting level, $n_{\rm opt}$, has been shown on each curve. Whenever the optimum nesting level changes as a function of $\beta$, the transition point has been highlighted on the graph with a large red oval. Up to this point, the optimum nesting level is the higher of the values given. From this point on, the lower given value becomes the optimum nesting level. The measurement error probability is set to $\delta=10^{-4}$. The coherence time of electron spins and nuclear spins are $\tau_{\rm e}=10$~ms and $\tau_{\rm n}=1$~s, respectively. }
\end{figure}

Based on above expressions, we compare the secret key generation rate, in the nominal mode of operation where no eavesdropper is present, for protocols 1 to 4. Our objective is to estimate the relevant parameters in \cref{eq:norm_skr} to get some insight into how these protocols are expected to perform in practice.

The nominal parameter values used in our numerical results are as follows. We fix $P_{S1}$ and $P_{S2}$ to 0.99 each, from which $T_1$ and $T_2$ can be calculated for each protocol. Note that $T_1$ and $T_2$ are increasing functions of $M$, which indirectly affects the waiting/decoherence time in our protocols. For our set of parameters, $T_1$ and $T_2$, respectively, end up to be somewhere between $1.5$ to $2$ times the average time that it takes to entangle all relevant memory pairs. We then calculate $\overline T_1$ and $\overline T_2$ to estimate the effect of decoherence on our system. We set the coherence time of electron and nuclear spins as $\tau_{\rm e}=10$~ms and $\tau_{\rm n}=1$~s, respectively, which is achievable in practice \cite{bernien2013heralded,bar2013solid,maurer2012room}. {The above values mostly reflect the back action on nuclear spins, when electron spins are being manipulated, but recent work with this type of memory \cite{abobeih2018one,kalb2018dephasing,bradley2019ten,pompili2021realization} has shown some progress with resolving this issue. We will therefore consider larger values of coherence time in our numerical results as well. The detector efficiency is set to $\eta_d=0.9$, which can be achieved by using superconducting single-photon detectors \cite{zhang2015advances}, offering negligible dark counts in our case. The distribution time for next-to-each-other electron-electron entanglement is set to $T_s= 5$~$\mu$s, which is on the same order of magnitude as the timing of internal operations reported in \cite{pompili2021realization}}. We choose the optical fiber as our channel with the speed of light being $c=2\times 10^5$ km/s and $L_{\rm att} = 22$~km.

Figure \ref{fig:SKR_beta} illustrates the performance of different protocols for generating secret keys as a function of {\sc cnot} gate error probability $\beta$, at electron-spin measurement error probability of $\delta=10^{-4}$, in the presence of depolarizing noise.  We have chosen several different values for the coupling efficiency, $\eta_c$, as well as three nominal distances of 100~km, 300~km, and 500~km. Such distances are perhaps too short to have an immediate impact in practice, but they are relevant to early demonstrations of quantum networks as being pursued in, e.g., Netherlands \cite{pompili2021realization}. For each value of $\beta$, we have found the optimum nesting level, {denoted by $n_{\rm opt}$ on each curve}, for each protocol, that maximizes the key rate in \cref{eq:norm_skr}. Figures \ref{fig:SKR_beta}(a)-(d) show system performance at different combinations of such parameters. Note that, for some parameter regimes, some protocols have not been able to generate a positive key rate, and, therefore, are absent from the relevant graph. We make several interesting observations from this figure, as summarized below:  
\begin{itemize}
    \item {\bf Observation 1:} Among different values chosen, in our simulation, for the total distance, Protocols 3 and 4 could only generate non-zero secret key rates at $L_{\rm tot}=100$~km. {For the chosen measurement error probability and coherence time parameters, even if we improve the coupling efficiency to $\eta_c=0.7$, there is still no key at $L_{\rm tot} \geq 200$~km for these two protocols.} This behavior is mainly because no distillation is considered in uncoded repeaters. {Given that conventional entanglement distillation techniques that do not rely on quantum error correction codes are probabilistic \cite{bennett1996purification, deutsch1996quantum}, it is not expected that they offer any improvement in key rate scaling either. The reason for this is that whenever we need to do a probabilistic operation, we need to repeat that until success. This requires additional classical communication to herald the success or failure of previous attempts, which results in additional delay and decoherence, both reducing the rate}. {This implies that at sufficiently short distances we may be better off not using any repeater node, and one can directly distribute an entangled state between the far-end users. This case is shown by the horizontal black solid line in Fig.~\ref{fig:SKR_beta}(d), and effectively represents the key rate at $n=0$ for protocol 3, given by
     \begin{align}
         R_0 = \frac{P_0(L_{\rm tot}) \times r_{\infty}^{\rm P3}}{2T_0}.
     \end{align}
     Because of the low number of operations needed in such a scenario, P3 at $n=0$ offers the best key rate for high values of $\beta$. At $L_{\rm tot} = 100$~km, if we restrict ourselves to $n\geq 1$, we observe that the highest secret key rates are generated by protocol 4 at low values of $\beta$ (hollow ovals in Fig. \ref{fig:SKR_beta}(d)). The distance that can be covered by such uncoded QRs is, however, limited. This observation implies that, without any encoding, the QR protocols may only be able to cover short distances, due to their low tolerance for errors.}
     %
    %
    \item {\bf Observation 2:} {Among the four protocols, protocol 2 seems to offer the best performance across a wide range of parameters. It even outperforms protocol 1, which, at the cost of higher computational overhead, is expected to have the best error correction/detection capabilities. One key reason for the superiority of protocol 2 in our numerical examples seems to be the use of freshly created entangled states for its entanglement distillation part, enabled here by error detection. To better understand this point, we need to compare the timing parameters for this protocol versus that of protocol 1. The first point is to note that $\overline T_1$ is lower for P2 than for P1. This is because in protocol 1 there are more memory pairs that need to be entangled, and that would make $T_1$, and consequently $\overline T_1$, longer for this protocol than protocol 2. For instance, at $L_{\rm tot} = 300$~km, $\eta_c = 0.5$, and $n=5$, $\overline T_1$ is roughly 0.0056~s and 0.0051~s for, respectively, P1 and P2. The second point is that, in the entanglement swapping stage, while $\overline T_2$ for protocol 2 is longer than that of protocol 1, it is typically much smaller than $\overline T_1$. For instance, at the same parameters as above, $\overline T_2$ is roughly 49~$\mu$s and 709~$\mu$s for, respectively, P1 and P2. This is because $\overline T_2$ corresponds to the entangling time for a single pair of memories, whereas $\overline T_1$ corresponds to entangling time for many pairs. This results in $\overline T_1 + \overline T_2$, which is the average time that nulcear spins have decohered before doing the BSM, to be comparable in the two protocols. In our example above, $\overline T_1 + \overline T_2$ is roughly 0.0056~s and 0.0058~s for, respectively, P1 and P2. Now, with this in mind, we can look at the situation when BSMs are performed in the two protocols. In P1, for error detection, we are using pairs of entangled links that have both decohered by roughly $\overline T_1 + \overline T_2$. In P2, however, one entangled link has roughly decohered by $\overline T_1 + \overline T_2$, whereas the other is freshly prepared and has only decohered for $T_0$. Given that $\overline T_1 + \overline T_2$ is almost the same in the two protocols, this asymmetry in the quality of entangled states used for distillation gives an edge to P2 over P1 and makes P2 more resistance, as compared to P1, against decoherence issues.  
    
    We should also bear in mind that the figure of merit that we use is normalized to the number of NV centers used, which, in the case of protocol 2, is almost half of that of protocol 1. This factor 2 is, however, compensated by $T_1+T_2$ term in \cref{eq:ent_generation_rate}, which, in the case of protocol 2 is less than two times that of protocol 1. In our example of $L_{\rm tot} = 300$~km, $\eta_c = 0.5$, and $n=5$, $T_1+T_2$ is 0.0067~s for P1 versus 0.0116~s for P2. The overall effect is then governed mainly by the decoherence time explained above, which is in favor of protocol 2.}
    \item {\bf Observation 3:} We notice that, for the optimum choice of nesting level, the inter-node distance varies roughly from 5-20~km. This, as we will see, will be a function of other relevant parameters such as coherence times and coupling efficiencies, and can slightly change either way in certain regimes. But, generally speaking, this inter-node distance is more manageable than that of third generation QRs, for which nodes are only apart by a few kms \cite{muralidharan2014ultrafast,borregaard2020one,azuma2015all,glaudell2016serialized,ewert2017ultrafast,lee2019fundamental}. It is, however, more demanding than that of probabilistic QRs, where the inter-node distance can be on the order of tens of kms \cite{piparo2013long,sangouard2011quantum}. {Another interesting observation is that, in some curves, the optimum nesting level goes down with $\beta$. This can be attributed to the fact that, at large values of $\beta$, it would be better to have fewer nodes so that the total number of operations, and the error that will be accumulated in the whole process, can better be managed.}
     
\end{itemize}

\begin{figure}[tb]
\includegraphics[width=\columnwidth]{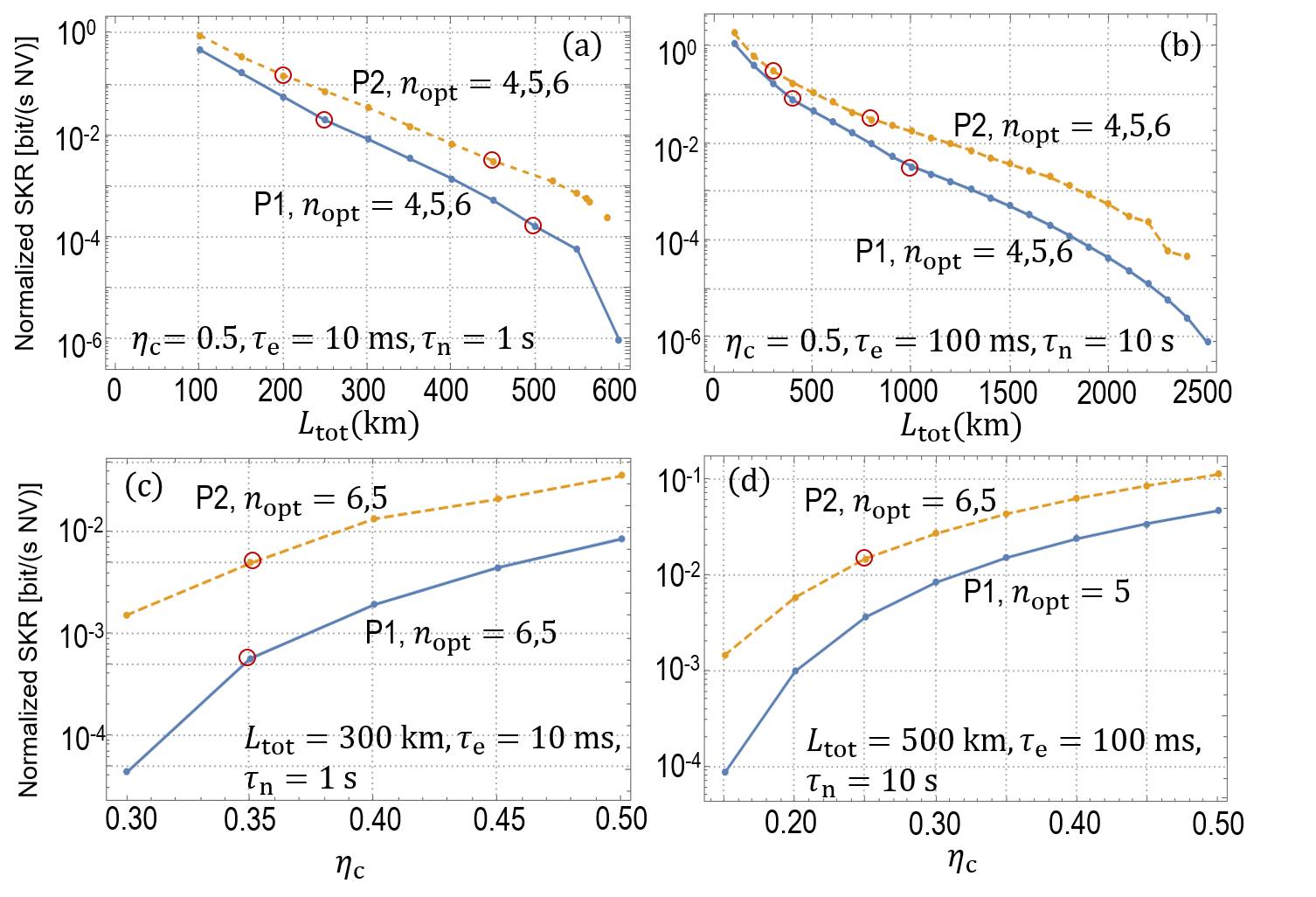}
\caption{\label{fig:other_variants} Comparison of normalized secret key rate as a function of total distance $L_{\rm tot}$ at (a) $\eta_c=0.5$, $\tau_{\rm e}=10$~ms, $\tau_{\rm n}=1$~s, (b) $\eta_c=0.5$, $\tau_{\rm e}=100$~ms, $\tau_{\rm n}=10$~s; and $\eta_c$ at (c) $L_{\rm tot} = 300$~km, $\tau_{\rm e}=10$~ms, $\tau_{\rm n}=1$~s, (d) $L_{\rm tot} = 500$~km, $\tau_{\rm e}=100$~ms, $\tau_{\rm n}=10$~s; for protocols 1 and 2. The {\sc cnot} gate error probability and measurement error probability are $\beta=10^{-3}$, $\delta=10^{-4}$, respectively. {Again, we have used large red ovals to indicate the point where the optimum nesting level, $n_{\rm opt}$, changes. In (a) and (b), the optimum nesting level at short distances is 4 and it goes up by one at red circles. In (c) and (d), it starts with the higher number and decreases by one at red circles.} }
\end{figure}

To further understand how protocols 1 and 2 compare to each other, in Fig.~\ref{fig:other_variants}, we investigate the sensitivity of these protocols to the total distance $L_{\rm tot}$, coupling efficiency $\eta_c$, and coherence times of nuclear, $\tau_{\rm n}$, and electron, $\tau_{\rm e}$, spins. Figures~\ref{fig:other_variants}(a) and (b) show the normalized key rate versus total distance for two different sets of coherence times, where $\tau_{\rm n}$ and $\tau_{\rm e}$ in Fig.~\ref{fig:other_variants}(b) are ten times that of Fig.~\ref{fig:other_variants}(a). {We observe that, in both figures, protocol 2 offers a higher key rate than protocol 1 for majority of distances. The two curves are closer at short distances as the corresponding values for $\overline T_1$ is too short for the coherence times considered. The two curves are even closer initially in Fig.~\ref{fig:other_variants}(b), where decoherence is less of an issue than in Fig.~\ref{fig:other_variants}(a). Figure~\ref{fig:other_variants}(b) also shows that such systems can cover distances in excess of 2000~km provided that coherence times are sufficiently long. At 2000~km, the optimum value for $L_0$ is over 30~km. The optimum nesting level increases with total distance as expected.}

{ The same tolerance to decoherence can be seen in Figs.~\ref{fig:other_variants}(c) and (d), where we compare the two protocols versus $\eta_c$. Now the two curves get closer at high values of $\eta_c$, which corresponds to shorter values for $\overline T_1$ and lower optimum nesting levels. For the range of parameters values considered in our numerical analysis, protocol 2 consistently offers higher key rates than protocol 1.}  

\begin{figure}[tb]
\includegraphics[width=\columnwidth]{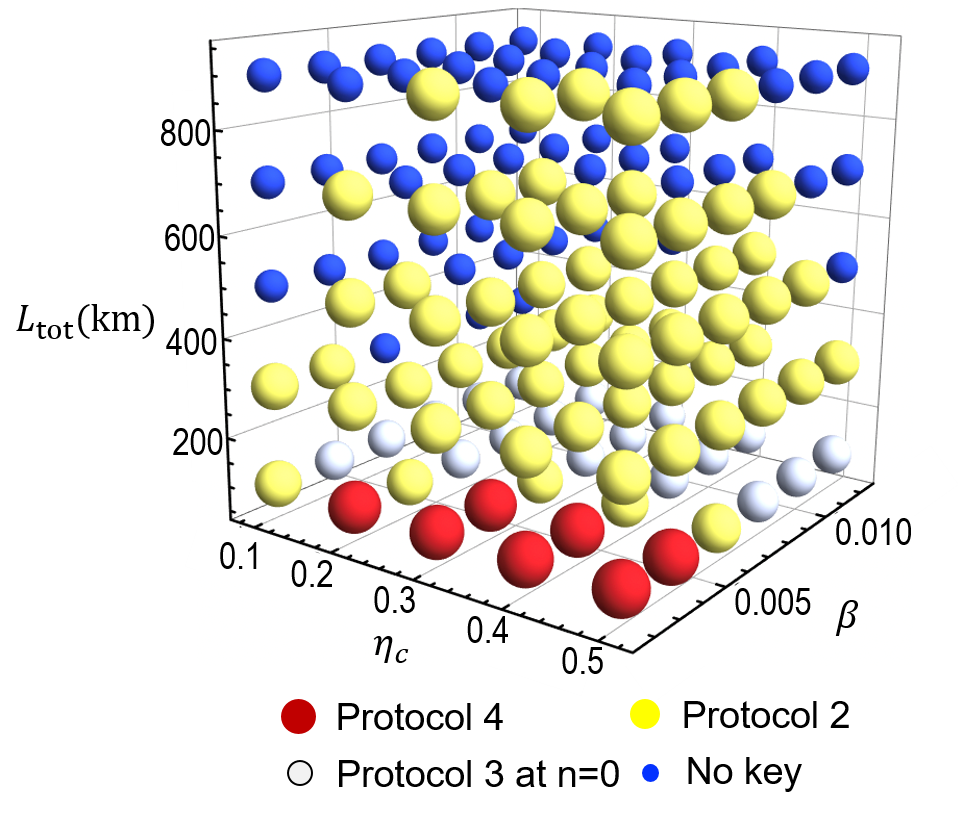}
\caption{\label{fig:3d} The region plot showing the distribution of the optimal QR protocol in a three-dimensional parameter space at $\delta=10^{-4}$, $\tau_{\rm e}=10$~ms, and $\tau_{\rm n}=10$~s.}
\end{figure}

Finally, in order to see, at different parameter regimes, which protocol would perform the best, in Fig.~\ref{fig:3d}, we have obtained the region plot highlighting the optimal QR structure that offers the highest key rate, at $\delta=10^{-4}$, $\tau_{\rm e}=10$~ms, $\tau_{\rm n}=10$~s, in a three-dimensional parameter space. We first note that, even with the improved nuclear coherence time, the uncoded QR protocols, i.e., protocols 3 and 4 still only work at $L_{\rm tot}=100$~km. At such short distance, these protocols could offer the best performance among all four. For longer distances and larger error probabilities, protocol 2 is most often the optimal choice. This leads to a practical conclusion that, for near-term implementations, the partially encoded QRs, which use fewer resources, might be the best option. 

\section{Conclusion}
\label{sec:conclusion}

In this work, we analyse two quantum repeater protocols with three-qubit repetition codes on NV centers in diamond platforms, with operation errors and decoherence noise being considered. We benchmark such encoded repeaters against uncoded structures by using QKD as a concrete application. We find that, the uncoded QRs only work and possibly offer some advantage at short distances. For longer distances, QRs with encoding are the optimal choice. {We notice that, for most practical regimes of interest, the protocol that relies on only partially encoded entangled states, hence consuming fewer physical resources, is the best-performing scheme.} This leads to a conjecture that, for the future implementation of encoded quantum repeaters, it seems that the partially encoded structures, rather than the fully encoded structure, are of more practical use. 

{ It is worth highlighting that, for any practical setup, there might be other parameters and processes that need to be properly modeled and accounted for in order to get a closer match to the experimental results. The work here offers some insight into and preliminary predictions for what we may expect in such experimental setups, but it does not replace the need for more accurate calculations when it comes to implementing such setups. Nevertheless, the results we have obtained here are promising in at least two aspects. First, they indicate that the required parameter regime for a decent operational quantum repeater setup is not far out of reach. Secondly, while the focus on experimental efforts today is mostly on simplest repeater structures, our work here shows that certain encoded structures should not be ruled out at this stage and can be part of our experimental agenda in the coming years.}

\appendix
\section{Derivation of decoherence parameters}
\label{sec:appendix}

The decoherence model in Eq.~(\ref{eq:depol_model}) for a single qubit system, where $d=2$ can be rewritten as
\begin{align}
  \mathcal{D}_{\rm depol} (\rho) =& \lambda_2 \rho + (1-\lambda_2) (\mathbb{I}_2 -\rho) \nonumber\\
  =& (2\lambda_2-1) \rho + (1-\lambda_2)\mathbb{I}_2 \nonumber\\
  =& (2\lambda_2-1) \rho + (1-\lambda_2) \frac{\rho + X\rho X +Y\rho Y +Z\rho Z}{2} \nonumber\\
  =&  \frac{3\lambda_2 -1}{2} \rho +\frac{1-\lambda_2}{2} (X\rho X +Y\rho Y +Z\rho Z)
\end{align}
where $X$, $Y$ and $Z$ are Pauli operations. For a two-qubit system, each qubit decoheres independently, which leads to 
\begin{widetext}
\begin{align}
    \mathcal{D}_{\rm depol}^{A} \circ \mathcal{D}_{\rm depol}^{B} (\rho_{AB})
   &=(\frac{3\lambda_2-1}{2})^2 \rho_{AB} + \frac{(3\lambda_2-1)(1-\lambda_2)}{4} [ (X_B,Y_B,Z_B) \rho_{AB} \left( \begin{array}{c} X_B \\ Y_B \\ Z_B \end{array} \right) + (X_A,Y_A,Z_A) \rho_{AB} \left( \begin{array}{c} X_A \\ Y_A \\ Z_A \end{array} \right) ] \nonumber\\
   &+ (\frac{1-\lambda_2}{2})^2  (X_A,Y_A,Z_A) (X_B,Y_B,Z_B)\rho_{AB} \left( \begin{array}{c} X_A \\ Y_A \\ Z_A \end{array} \right)\left( \begin{array}{c} X_B \\ Y_B \\ Z_B \end{array} \right).
\label{eqAppen:depol_2qubit}
\end{align}
\end{widetext}
If the state $\rho_{AB}$ is a Bell diagonal state, we can verify that the output state obtained from \eqref{eqAppen:depol_2qubit} is equivalent to \cref{eq:depol_model2}. Note that, in \cref{eqAppen:depol_2qubit}, a Bell state remains intact by the following operators: $\mathbb{I}_A\mathbb{I}_B, X_A X_B, Y_AY_B, Z_AZ_B$. Therefore, the fidelity of the output state with respect to the input Bell state is given by the sum of the corresponding coefficients in \eqref{eqAppen:depol_2qubit}: 
\begin{align}
 \lambda_4=\frac{1}{4} (3\lambda_2 -1)^2 + \frac{3}{4} (1-\lambda_2)^2,   
\end{align}
which is the same as \cref{eq:lambda4}.

To obtain the decoherence effect on a six-qubit system, we have to apply the single qubit depolarizing model in \cref{eq:depol_model} on each qubit independently, and calculate the tandem effect. This results in a lengthy expression for the output state, which we will not reproduce here. But, it can be verified that the operators that map the encoded Bell state $|\tilde{\Phi}^+\rangle_{\mathbf{AB}}$ to itself are given by: $\mathbb{I}^{\otimes6}$, $X^{\otimes6}$, $Y^{\otimes6}$, $Z^{\otimes6}$, $X^{\otimes2} Y^{\otimes4}$, $X^{\otimes4} Y^{\otimes2}$, $Z^{\otimes4} \mathbb{I}^{\otimes2}$ and $Z^{\otimes2} \mathbb{I}^{\otimes4}$. Again, one can calculate the corresponding fidelity for the output state, with respect to $|\tilde{\Phi}^+\rangle_{\mathbf{AB}}$, by accounting for the coefficients of the relevant terms to obtain
\begin{align}
    \lambda_{64}= &\frac{1}{64} [(3\lambda_2-1)^6 +33 (1-\lambda_2)^6 \nonumber\\ &+15(3\lambda_2-1)^2(1-\lambda_2))^4 \nonumber\\
    &+ 15(3\lambda_2-1)^4(1-\lambda_2)^2],
\end{align}
which is equivalent to \cref{eq:lambda64}.

\begin{acknowledgments}
This project is funded by the European Union's Horizon 2020 research and innovation programme under the Marie Sklodowska-Curie grant agreement number 675662 (QCALL) and UK EPSRC Grant EP/M013472/1. The data generated by this work can be reproduced by the methodology and equations given in the manuscript.
\end{acknowledgments}

%

\end{document}